\documentclass[prc,superscriptaddress,unsortedaddress,twocolumn]{revtex4}
\usepackage{graphicx,epsf,amsmath,color}
\def\br{\mbox{\boldmath $r$}}

\def\bfsigma{\mbox{\boldmath $\sigma$}}

\begin{document}
%\begin{CJK*}{SJIS}{}
%\draft
\title{$\Xi$ hyper-nuclear states predicted by NLO chiral baryon-baryon
interactions}
%\author{M. Kohno (‰Í–ì'ʘY)}
\author{M. Kohno}
%\email[]{kohno@rcnp.osaka-u.ac.jp}
\affiliation{Research Center for Nuclear Physics, Osaka University, Ibaraki 567-0047,
Japan\\}

\author{K. Miyagawa}
\affiliation{Graduate School of Science, Okayama University of Science,
Okayama 700-0005, Japan\\}
\affiliation{Research Center for Nuclear Physics, Osaka University, Ibaraki 567-0047,
Japan\\}

\begin{abstract}
The $\Xi$ single-particle potential obtained in nuclear matter with the next-to-leading
order baryon-baryon interactions in chiral effective field theory is applied to
finite nuclei by an improved local-density approximation method. As a premise, phase
shifts of $\Xi N$ elastic scattering and the results of Faddeev calculations for the
$\Xi NN$ bound state problem are presented to show the properties of the $\Xi N$
interactions in the present parametrization. First, the $\Xi$ states in $^{14}$N are
revisited because of the recent experimental progress, including the discussion
on the $\Xi N$ spin-orbit interaction that is relevant to the location of
the $p$-state. Then the $\Xi$ levels in $^{56}$Fe are calculated. In particular,
the level shift which is expected to be measured experimentally in the near future
is predicted. The smallness of the imaginary part of the $\Xi$ single-particle
potential is explicitly demonstrated.
\end{abstract}
%\pacs{??}

\maketitle
%\end{CJK*}

\section{Introduction}
New experimental information on the $\Xi$-nucleus interaction is increasing
from the analyses of experiments at J-PARC. The first observation of
twin single-$\Lambda$ hypernuclei in the experiment at J-PARC identified
a $\Xi^-$-$^{14}$N bound state with the binding energy $B_\Xi=1.27\pm0.21$
MeV \cite{HAY21}. The energy is close to the candidate of the $\Xi^-$-$^{14}$N
state observed in the previous KEK E373 experiment \cite{NAK15} with
$B_\Xi=1.03\pm0.18$ MeV. 
In the near future, further observation
of $\Xi$ bound states in nuclei is expected. The inclusive spectra of $(K^-,K^+)$
reactions on nuclei \cite{NAG18} should provide the properties of $\Xi$-nucleus
potential in a wide energy range. Another ongoing experiment to detect $\Xi$
atomic level shifts by measuring electromagnetic transition spectra \cite{JPE03}
is also valuable to inform the $\Xi$-nucleus potential in the surface region.   

On the theoretical side, the construction of baryon-baryon interactions in the
strangeness $S=-2$ sector has been developed in the framework of chiral effective
field theory (ChEFT) \cite{Pol07,HAID16,HAID19}.
The lattice QCD method by the HAL-QCD
group also provides the parametrization of the $S=-2$ interactions \cite{Ino19,Sas20}.
Both descriptions are based on the QCD, namely the underlying theory of hadrons
and their interactions, either in a direct way or by way of low-energy
chiral effective field theory. $\Xi N$ interactions of these two methods are,
interestingly, are resembling even at the quantitative level, as is shown in the
following section. The $S=-2$ sector of the octet baryon-baryon interactions
is the middle of the possible strangeness contents from $S=0$ to $-4$ and
therefore all the combinations of the flavor SU(3) bases participate in the
feature of the interactions. 

Because it is not feasible in the near future to measure directly $\Xi$-nucleon
scattering, the information on the $\Xi$ bound states is the chief source for
the $\Xi N$ interactions. It is hard, however, to find detailed spin and isospin
structure of the $\Xi N$ interaction from the analysis of the bound state data
in itself. It is necessary to compare the experimental data with the results
of microscopic calculations using theoretical interactions as reliable as possible.    

One of the present authors reported, in Ref. \cite{MK19},
the properties of the $\Xi$-nucleus single-particle potential which are
obtained on the basis of $G$-matrix calculations in symmetric nuclear matter
with next-to-leading order (NLO) ChEFT potentials constructed by the
J\"{u}lich-Bonn-M\"{u}nchen group \cite{HAID16,HAID19}. There, $\Xi$
potentials in light nuclei such as $^9$Be, $^{12}$C, and $^{14}$N are predicted through
the translation of the potential in infinite matter to that in a finite nucleus
by an improved local-density approximation (ILDA) method. In view of the current
experimental efforts, it is meaningful to revisit the case of $^{14}$N and extend
the calculation of the chiral $\Xi$ potential to heavy nuclei such as $^{56}$Fe.

In Sec. 2, the basic properties of the chiral NLO $\Xi N$ interactions are elucidated
by  presenting $\Xi N$ phase shifts, and the results of the Faddeev calculations
for searching a $\Xi NN$ bound state. The $\Xi^-$ single-particle potentials in
heavier nuclei are studied in Sec. III. First, the $\Xi^-$ states predicted on $^{14}$N
are revisited. The probable $0p$ $\Xi^-$ state experimentally observed \cite{HAY21}
is conducive to the discussion of the $\Xi$ spin-orbit single-particle potential.
Next, the potential for $^{56}$Fe is presented, for which the measurement of
the level shift of a certain atomic level is aimed in the J-PARC experiment.
The very small imaginary part of the $\Xi$ single-particle potential is demonstrated.
Summary follows in Sec. IV.

\section{Properties of chiral NLO $\Xi N$ interaction}
\subsection{$\Xi N$ phase shift}
\begin{figure}[t]
\centering
 \includegraphics[width=0.45\textwidth,pagebox=cropbox,clip]{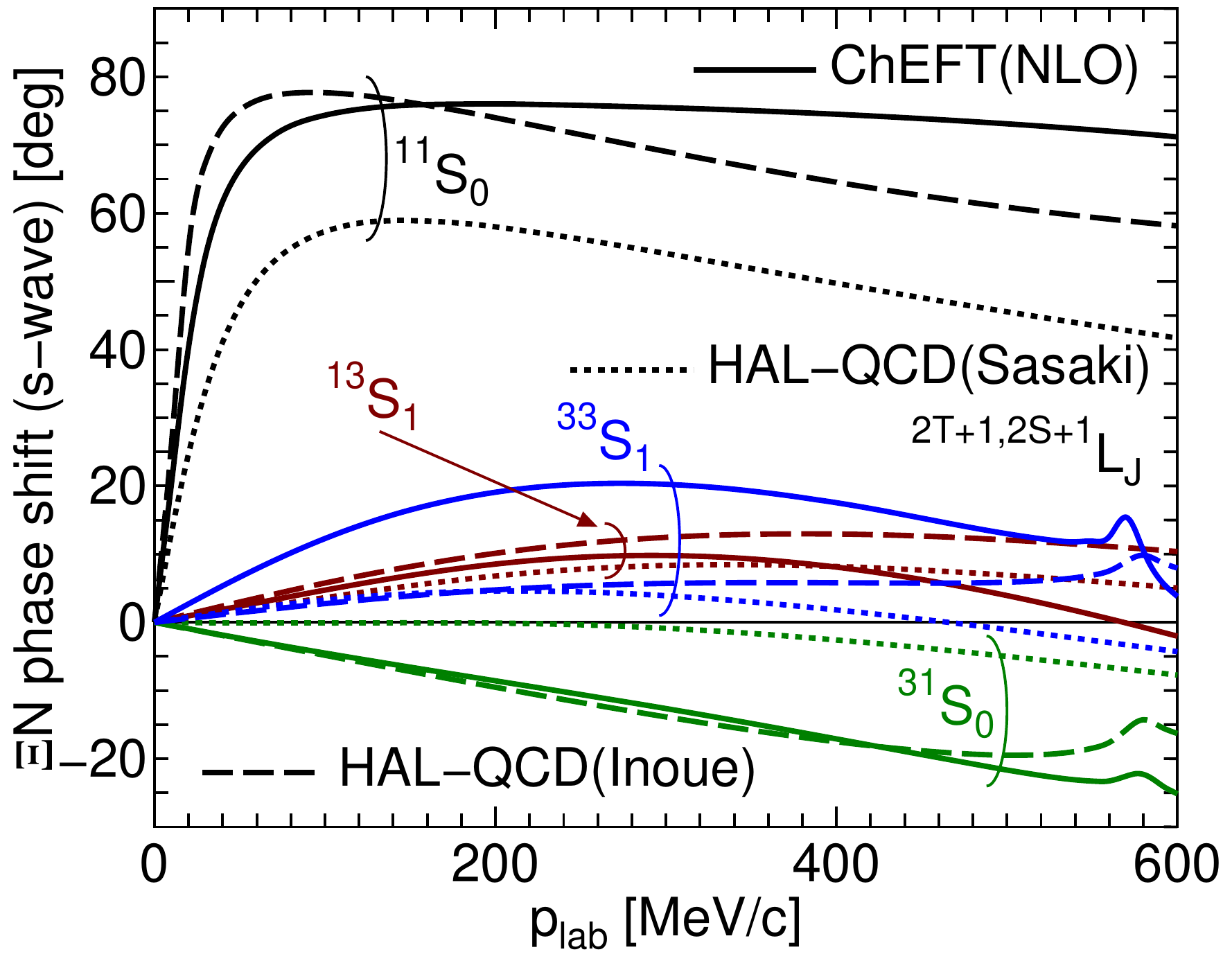}
\caption{
$\Xi N$ $s$-wave phase shifts calculated with NLO ChEFT interactions
are shown by solid curves
with the notation $^{2T+1,2S+1}L_J$ for specifying the spin $S$ and isospin $T$
channel. Phase shifts
with two sets of the parametrization based on the HAL-QCD calculations are
also shown: one is the full parametrization by Inoue et al. \cite{Ino19} (dashed)
and the other is the fit for $t/a=12$ by
Sasaki et al. \cite{Sas20} (dotted).
}
\label{fig:xnsph}
\end{figure}

It is basic to evaluate phase shifts of elastic scattering to elucidate the properties
of the $\Xi N$ interaction in each spin and isospin channel. The $s$-wave phase shifts
calculated with an updated version of the chiral NLO interactions \cite{HAID19} are
shown in Fig. \ref{fig:xnsph} by the solid curves. Calculations are in the isospin base. 
That is, the average masses are assigned for the N, $\Sigma$, and $\Xi$
baryons, respectively.
The phase shifts with the interactions parametrized
on the basis of HAL-QCD calculations are also included for comparison.
There are two sets of parametrization based on the same HAL-QCD
calculations. The dashed curves represent the results of the potential by
Inoue \textit{et al.} \cite{Ino19} in which the baryon-channel coupling
components are explicitly parametrized as a local function.
The dotted curves are the results of
the potential of the fit for $t/a=12$
by Sasaki \textit{et al.} \cite{Sas20} in which
the effects of the tensor coupling and the baryon-channel coupling except for
$\Lambda\Lambda$ are simulated
by a local $\Xi N$ potential in coordinate space.

It is seen that three potentials predict qualitatively similar behavior of
the phase shifts in all spin and isospin states. The interaction in the isospin $T=1$
and $^1S_0$ state is repulsive, and the interactions in the remaining
three states are attractive. Nevertheless, some quantitative differences are
remarked. The repulsion of the $T=1$ $^1$S$_0$ part of the Sasaki potential is very weak.
The attraction in the $T=1$ $^3$S$_1$ state of the HAL-QCD parametrization is
smaller than that of ChEFT. The attraction
in the $T=0$ $^3S_1$ state, in which no baryon-channel coupling is present,  is weak.
The $T=0$ $^1S_0$ state is most attractive, although no bound state exists.
This attraction originates from the coupling to the $\Lambda\Lambda$
as well as $\Sigma\Sigma$ states, though the effect of the $\Xi N$-$\Lambda\Lambda$
coupling is smaller than that of the $\Xi N$-$\Sigma\Sigma$ coupling in the
HAL-QCD potentials.
The attractive character in the $T=1$ $^3S_1$ state is not so prominent as in
the $T=0$ $^1S_0$ state but plays an important role to generate an attractive $\Xi$
single-particle potential in a nucleus because of the spin-isospin weight factor
of $(2S+1)(2T+1)=9$.

\begin{figure}[t]
\centering
 \includegraphics[width=0.45\textwidth]{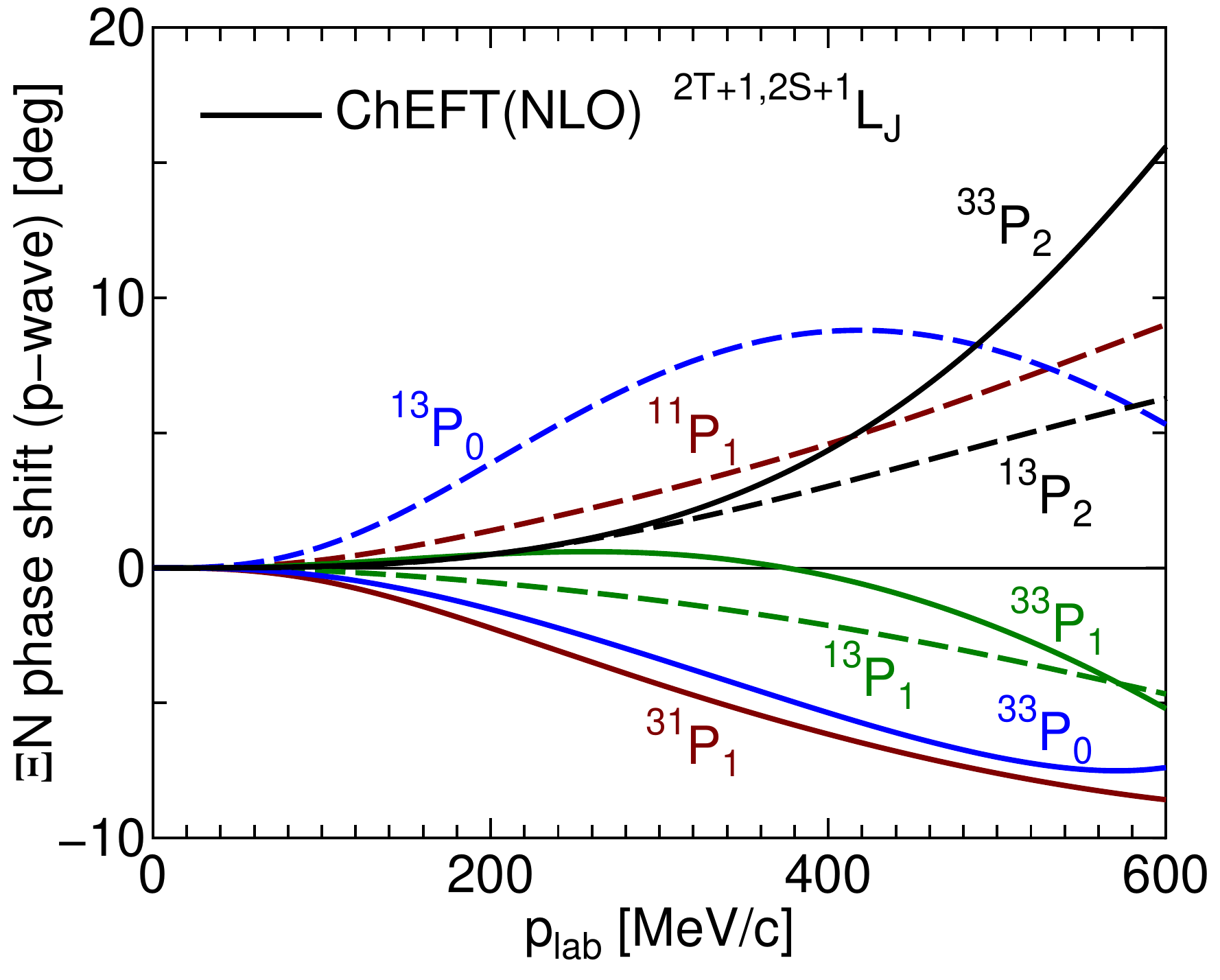}
\caption{
$\Xi N$ $p$-wave phase shifts calculated with NLO ChEFT interactions
are shown. Those of the isospin
$T=0$ ($T=1$) states are denoted by dashed (solid) curves.
}
\label{fig:xnpph}
\end{figure}

The uncertainties in the ChEFT parametrization of $p$-waves are
larger than those in the $s$-waves \cite{HAID16}. The anti-symmetric spin-orbit
interactions, which couples the spin-single and triplet states with the same total
spin $J$, are absent in the present chiral NLO interactions \cite{HAID16}.
Still, it is meaningful to present $p$-wave phase shifts for inferring the
effects of the $p$-waves on the $\Xi$-nucleus potential. 
The $p$-wave phase shifts calculated with the chiral NLO interactions \cite{HAID16}
are shown in Fig. \ref{fig:xnpph}. The phase shifts are rather small, except for in the
$T=1$ $^3$P$_2$ state, the attraction of which grows with increasing energy.
The corresponding attractive contribution to the $\Xi$ single-particle potential
in symmetric nuclear matter was presented in Fig. 2 of Ref. \cite{MK19}.
It is also seen in that figure that the contributions to the $\Xi$ single-particle
from other $p$-states are small and tend to cancel each other among them. 

\subsection{$\Xi NN$ three-body system}
It is important to figure out whether the chiral NLO interaction can support a
$\Xi NN$ three-body bound state. The results of the Faddeev calculation
for the $\Xi NN$ bound state problem are recapitulated in this section.
In the present calculations, two-body $\Xi N$ $T$-matrices are first prepared by
solving a baryon-channel coupled Lippmann-Schwinger equation in momentum space.
In the isospin $T=0$ case, the $\Xi N$-$\Lambda\Lambda$-$\Sigma\Sigma$
coupling is present in the $^1$S$_0$ state, while no baryon-channel coupling exists
in the $^3$S$_1$-$^3$D$_1$ tensor correlated state. In the $T=1$ case, the
$\Xi N$-$\Lambda\Sigma$ coupling is present in the $^1$S$_0$ state, and
the $\Xi N$-$\Lambda\Sigma$-$\Sigma\Sigma$ coupling takes place
in the $^3$S$_1$-$^3$D$_1$ tensor correlated state.
Then, the evaluated $T$-matrices are used in the Faddeev equation for
the $\Xi NN$ bound-state problem:
\begin{align}
 \Psi^{(23)}=& G_0 T_{NN} (1-P_{23}) \Psi^{(12)}, \\
 \Psi^{(12)}=& G_0 T_{\Xi N} (\Psi^{(23)}-P_{23} \Psi^{(12)}),
\end{align}
where $G_0$ is a three-particle propagator, $\Psi^{(ij)}$ is the Faddeev component,
and $P_{23}$ is the transposition operator for the 2-3 pair with assigning
the number 1 to $\Xi$ and the remaining 2 and 3 to the nucleons. 
This procedure means that while the pairwise
correlation is fully solved, the entire three-baryon coupling is not considered.
The interactions are also restricted to the $s$-wave except for the
tensor-coupled $d$-wave. The Coulomb force is also not taken into account.
Still, the calculation is an important attempt for a realistic description of the
$\Xi NN$ system. 
 
In the literature, possible $\Xi NN$ bound states have been reported \cite{GV16,Fil17},
using the $s$-wave single-channel $\Xi N$ potential simulating the Nijmegen ESC08c
model \cite{ESC08} for the $\Xi N$ interaction and the central  $s$-wave
Malfliet-Tjon $NN$ potential \cite{MTNN} for the $NN$ interaction.
The Faddeev calculations in Ref. \cite{Fil17} without the Coulomb force show
that the bound state exists at the binding energy $B= 17.2$ MeV in the
spin-isospin $(S,T)=(3/2,1/2)$ state and $B=2.9$ MeV in the $(S,T)=(1/2,3/2)$
state. These results are reproduced in our momentum-space Faddeev
calculations.
It is noted, however, that a substantial revision was made for ESC08c to construct the
new version as ESC16 by those authors \cite{ESC16}.

The situation is different when the chiral NLO $S=-2$ interactions are used
together with the N$^3$LO $NN$ interactions \cite{EGM05}. The results of our
Faddeev calculations show that no hyper-nuclear bound $\Xi NN$ system is
expected in every possible spin-isospin channel. It is also ascertained
that even if the repulsive T=1 $^1$S$_0$ $\Xi N$ interaction is omitted,
the $\Xi NN$ system is not bound.
The details are reported in a separate paper \cite{MIY21}.

\section{$\Xi$ bound states in finite nuclei}
Although the $\Xi NN$ system is not bound with the present NLO chiral
interactions, the $\Xi$ hyperon can be bound in heavier nuclei due
to the attraction in the $T=1$ $^3$S$_1$ channel with the statistical factor
of $(2S+1)(2T+1)$. As shown in Ref. \cite{MK19}, the $\Xi^-$ single-particle
potentials predicted for $^9$Be, $^{12}$C, and $^{14}$N by the ILDA method
using the $G$-matrices in symmetric nuclear matter with the NLO chiral interactions
are rather shallow but enough attractive to support hyper-nuclear bound states.
In this section, first, the calculated $\Xi^-$-$^{14}$N bound states are revisited
concerning the recent experimental data and the possible effect
of the $\Xi$ spin-orbit potential for the $p$-state.
Next, $\Xi^-$ bound states in $^{56}$Fe are presented. In particular,
the atomic level shift of $\Xi^-$ in $^{56}$Fe is focused, for which the X-ray
spectroscopy experiment to detect it ongoing at J-PARC \cite{JPE03}.
The very small imaginary part of the $\Xi$ potential is exemplified,
which was not included in Ref. \cite{MK19}.

\subsection{$\Xi^-$-$^{14}$N bound states and $\Xi$ spin-orbit potential}
After the prediction for $\Xi^-$ bound states in $^{14}$N was reported
in Ref. \cite{MK19} based on the NLO ChEFT $S=-2$ interactions, new
experimental information \cite{HAY21} was obtained through the first observation of
twin single-$\Lambda$ hypernuclei: $\Xi^- +^{14}\mbox{N}\rightarrow ^{10}_\Lambda
\mbox{Be}+^5_\Lambda\mbox{He}$. The state is a probably $0p$ level at
$1.27\pm 0.21$ MeV. Further observation of $\Xi^-$ states in $^{14}$N was
also reported \cite{YOS21} from the analyses of the data of KEK and J-PARC
experiments. That is, three candidates for the $0s$ state are at $8.00\pm 0.77$,
$4.96\pm 0.77$, and $6.27\pm 0.27$ MeV, respectively, and a possible
$0p$ state is at $0.90\pm 0.62$ MeV. These energies are shown
on the left side of Fig. \ref{fig:xin14e}.

\begin{figure}[b]
\centering
 \includegraphics[width=0.45\textwidth]{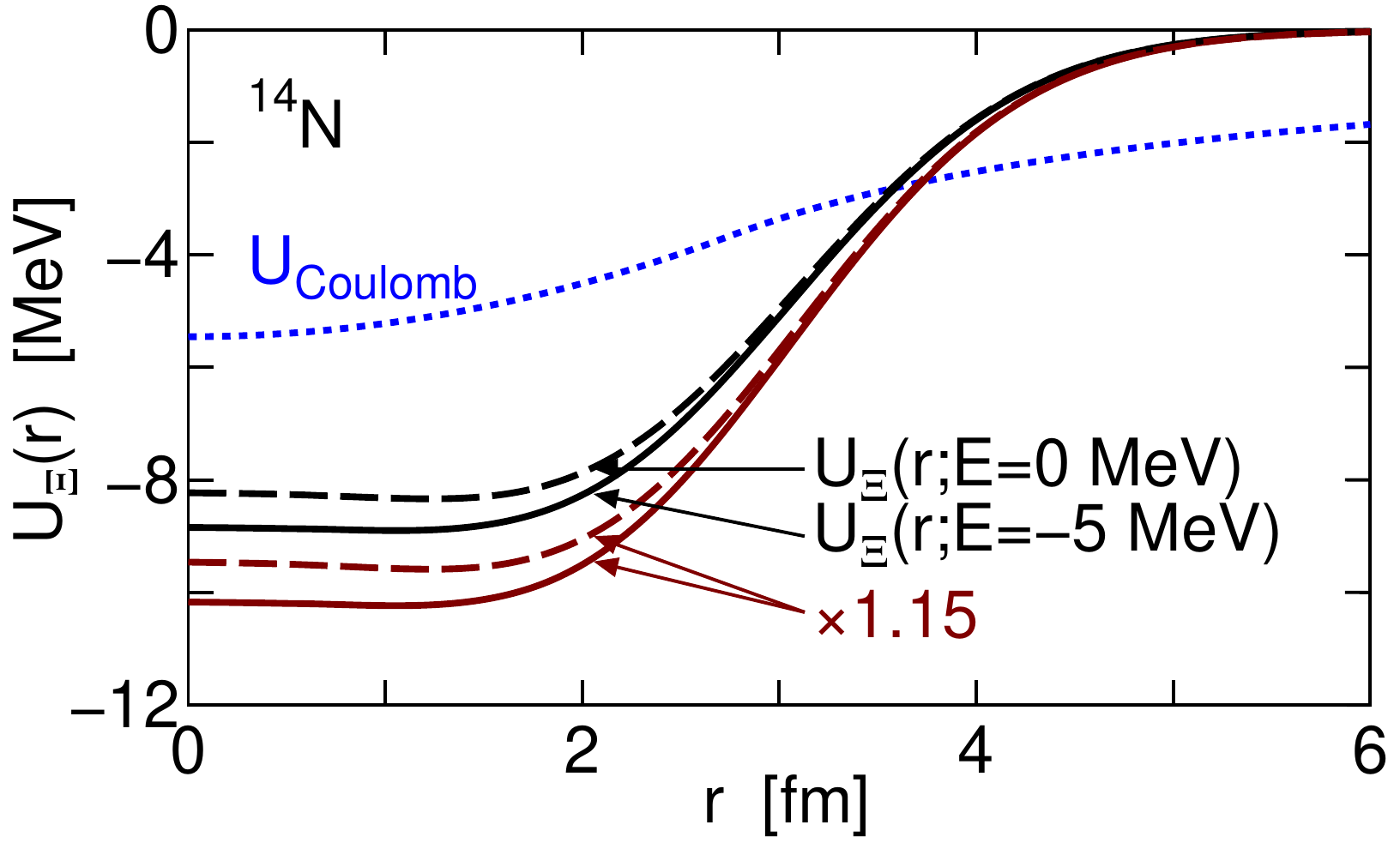}
\caption{
Energy-dependent $\Xi$-$^{14}$N single-particle potential by the ILDA
method, using $\Xi N$ $G$-matrices in symmetric nuclear matter with the
NLO ChEFT interactions \cite{HAID19}.
The potentials enhanced by a factor of 1.15 are also shown.
$U_{\mbox{Coulomb}}$ is the potential  of uniform charge distribution with a radius
of $R_C=1.15 A^{1/3}$ fm.
}
\label{fig:xin14p}
\end{figure}

\begin{figure}[t]
\centering
 \includegraphics[width=0.45\textwidth]{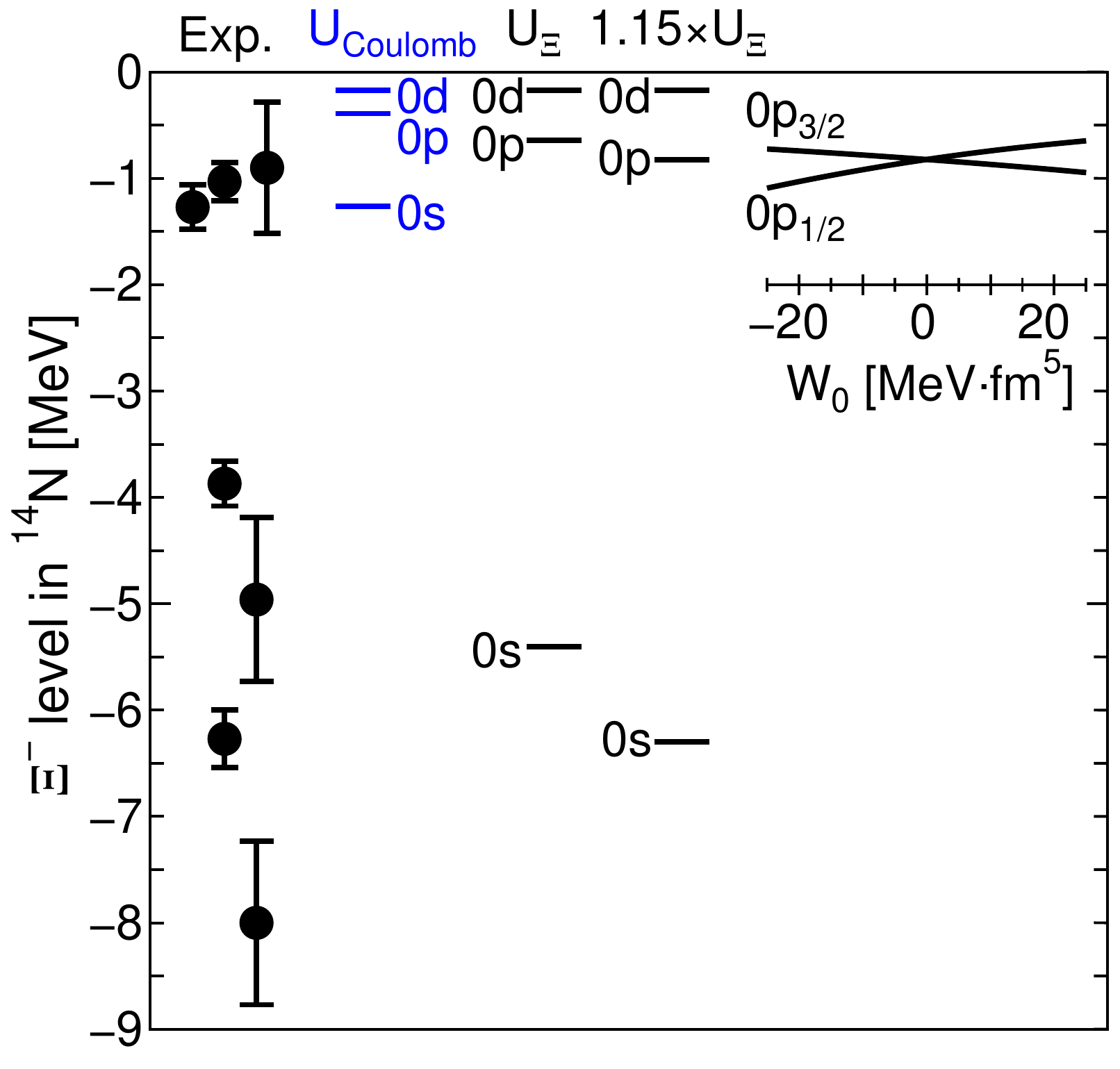}
\caption{
$\Xi^-$ single-particle states in $^{14}$N. Experimental data on the left side
are taken from the compilation in Ref. \cite{YOS21}. The Coulomb attraction is treated
by the potential of uniform charge distribution with a radius of $R_c=1.15 A^{1/3}$ fm.
The ILDA potential is energy-dependent. $U_\Xi(r;E=-5\mbox{MeV})$
is employed for the $0s$ state and $U_\Xi(r;E=0\mbox{MeV})$ for the $0p$ and $0d$ states.
It is demonstrated that an enhancement factor of 1.15 is needed to fit the $0s$ energy
at 6.27 MeV. The shifts of the $0p_{3/2}$ and $0p_{1/2}$ energies due to the
addition of the $\Xi N$ spin-orbit potential is shown on the right side, as a function
of the strength $W_0$ of Eq. \ref{eq:w}.
}
\label{fig:xin14e}
\end{figure}

Observing that the calculated energies in Ref. \cite{MK19} correspond
reasonably well to these experimental data, additional calculations are given
in this subsection. First, if the experimental energy of $6.27\pm 0.27$ MeV
is taken seriously, the calculated $\Xi^-$ $0s$ energy of $-5.40$ MeV in
Table II in Ref. \cite{MK19} is slightly short. To reproduce the range of
the empirical energy, $-6.00\sim -6.54$ MeV, it is needed to multiply
a factor of $1.10\sim1.19$ to the calculated $\Xi$-$^{14}$N potential.
This factor appears within the uncertainties of the $G$-matrix calculations
and the ILDA method. The calculated $\Xi$-$^{14}$N single-particle potential
is shown in Fig. \ref{fig:xin14p}. The potential by the ILDA method
is energy-dependent. $U_\Xi(r;E=-5\mbox{MeV})$ is employed for the $0s$ state,
and $U_\Xi(r;E=0\mbox{MeV})$ is for the $0p$ and $0d$ states.
The evaluated single-particle energies are presented in Fig. \ref{fig:xin14e}
both for $U_\Xi$ and $1.15\times U_\Xi$. The position of
the $0p$ level is reasonable. It is noted that the Coulomb
0d state is hardly affected by the addition of the hyper-nuclear potential
$U_\Xi(r;E=0\mbox{MeV})$.

Another subject to discuss here is the effect of a $\Xi$-nucleus spin-orbit
potential. If the $\Xi$ spin-orbit interaction is not negligible, the location
of the $0p$ state does not simply imply the strength of the $\Xi$ central
single-particle potential.
Although the ground state of $^{14}$N is not simply shell-closed, it is
instructive to estimate how the $\Xi$ $0p$ level in $^{14}$N is
affected by the possible spin-orbit potential in a mean-field consideration;
that is, without considering the detailed structure of the $^{14}$N ground state.

The interesting feature of the $\Xi$ spin-orbit single-particle potential in nuclei
is that the potential may be repulsive in contrast to the attractive nucleon
spin-orbit potential. Various theoretical studies have predicted a repulsive
$\Xi$ spin-orbit mean-field, though the strength is considerably smaller than
that of the nucleon. When the spin-orbit potential is repulsive, the downward
shift of the single-particle level with $j_{<}=\ell-\frac{1}{2}$ is twice as large as
that of the level with $j_{>}=\ell+\frac{1}{2}$ for the attractive one.

In a relativistic mean-field description \cite{CW91,TSHT98,CM20}, the repulsive
spin-orbit mean-field is brought about by an $\omega$-meson exchange with
the tensor coupling. The repulsive character is also predicted in
a microscopic description based on two-body $\Xi N$ interactions
constructed in a non-relativistic SU(6) quark model \cite{Fuji00}, in which
the contribution from the ordinary spin-orbit component of the $\Xi N$ interaction
is attractive, while the anti-symmetric spin-orbit component contributes opposite
and the net spin-orbit single-particle potential becomes repulsive. In all these
estimations, the repulsive strength is one-fifth of the attractive strength of
the nucleon spin-orbit potential or less.

The effective spin-orbit strength generated by the bare baryon-baryon interaction
is properly measured by the Scheerbaum factor \cite{SCH76} calculated in nuclear matter.
The expression for the nucleon case was extended to the hyperon cases in Ref. \cite{Fuji00}.
The Scheerbaum factor for $\Xi$ in symmetric nuclear matter with the Fermi
momentum $k_F$ reads
\begin{eqnarray}
 S_\Xi (\bar{q})=\frac{\zeta(1+\zeta)^2}{8k_F^3}\sum_{JT} (2J+1)(2T+1)
\int_0^{q_{max}} dq  \nonumber \\
 \times W(\bar{q},q)\{ (J+2)G_{1J+1,1J+1}^{JT}(q)+G_{1J,1J}^{JT}(q) \nonumber \\
 -(J-1)G_{1J-1,1J-1}^{JT}(q) \}.
\label{eq:lss}
\end{eqnarray}
Here, $\zeta\equiv m_N/m_\Xi$, $q_{max}=\frac{1}{2}(k_F+\bar{q})$ and
the weight factor $W(\bar{q},q)$ is defined by
\begin{equation}
 W(\bar{q},q)=\left\{ \begin{array}{l}
 \theta (k_F-\bar{q})\;\;\mbox{for} \;\; 0\leq q\leq \frac{|k_F-\bar{q}|}{2} \\
 \frac{k_F^2-(\bar{q}-2q)^2}{8\bar{q}q}\;\;\mbox{for}\;\;
 \frac{|k_F-\bar{q}|}{2} \leq q\leq \frac{k_F+\bar{q}}{2},
 \end{array} \right.
\end{equation}
where $\theta (k_F-\bar{q})$ is a step function.
In Eq. \ref{eq:lss}, $G_{1\ell',1\ell}^{JT}$ is the abbreviation of
the momentum-space diagonal $G$-matrix element in the spin-triplet
channel with the total spin $J$ and total isospin $T$.

The above definition of $S_\Xi$ is different from the original constant
in Ref. \cite{SCH76} by a factor of $-\frac{2\pi}{3}$. Then, $S_\Xi$ can be
identified with the strength $W_0$ of the $\delta$-type effective two-body
spin-orbit interaction customarily used in Skyrme-Hartree-Fock
calculations \cite{SKY}:
\begin{equation}
 i W_0(\bfsigma_1+\bfsigma_2)\cdot [\nabla_r \times \delta(\br)\nabla_r].
\label{eq:w}
\end{equation}

In the present NLO ChEFT interaction, the antisymmetric spin-orbit term is
not included by putting the pertinent low-energy constant to zero.
Therefore, the spin-orbit interaction in the nuclear medium is expected to be
attractive. The actual $G$-matrix calculation in symmetric nuclear matter gives
$S_\Xi \simeq 21.5$ MeV$\cdot$fm$^5$ at $\bar{q}\approx 0.7$ fm$^{-1}$
that is the value prescribed by  Scheerbaum on the basis of the wavelength
of the density distribution.
This value is about one-fifth of $S_N=102$ MeV$\cdot$fm$^5$ with the same sign
\cite{MK13}. It is noted that $S_N=102$ MeV$\cdot$fm$^5$ is somewhat
smaller than the typical value of $S_N=120$ MeV$\cdot$fm$^5$ used in
Skyrme-Hartree-Fock calculations in the literature.

Changes of the $\Xi^-$ $0p$ level in $^{14}$N of the potential
$1.15\times U_\Xi(E=0 MeV)$ depending on the spin-orbit parameter $W_0$
from $-25$ to 25 MeV$\cdot$fm$^5$ are depicted in Fig. \ref{fig:xin14e}. The
negative sign of $W_0$ means the repulsive spin-orbit potential. If the spin-orbit
potential strength is about one-fifth of that of the nucleon, the energy splitting
of the shallow $0p_{1/2}$ and $0p_{3/2}$ states in $^{14}$N is about 0.4 MeV.

\subsection{$\Xi$ bound states in $^{56}$F$\mbox{e}$ and atomic level shifts}
\begin{figure}[t]
\centering
 \includegraphics[width=0.45\textwidth,pagebox=cropbox,clip]{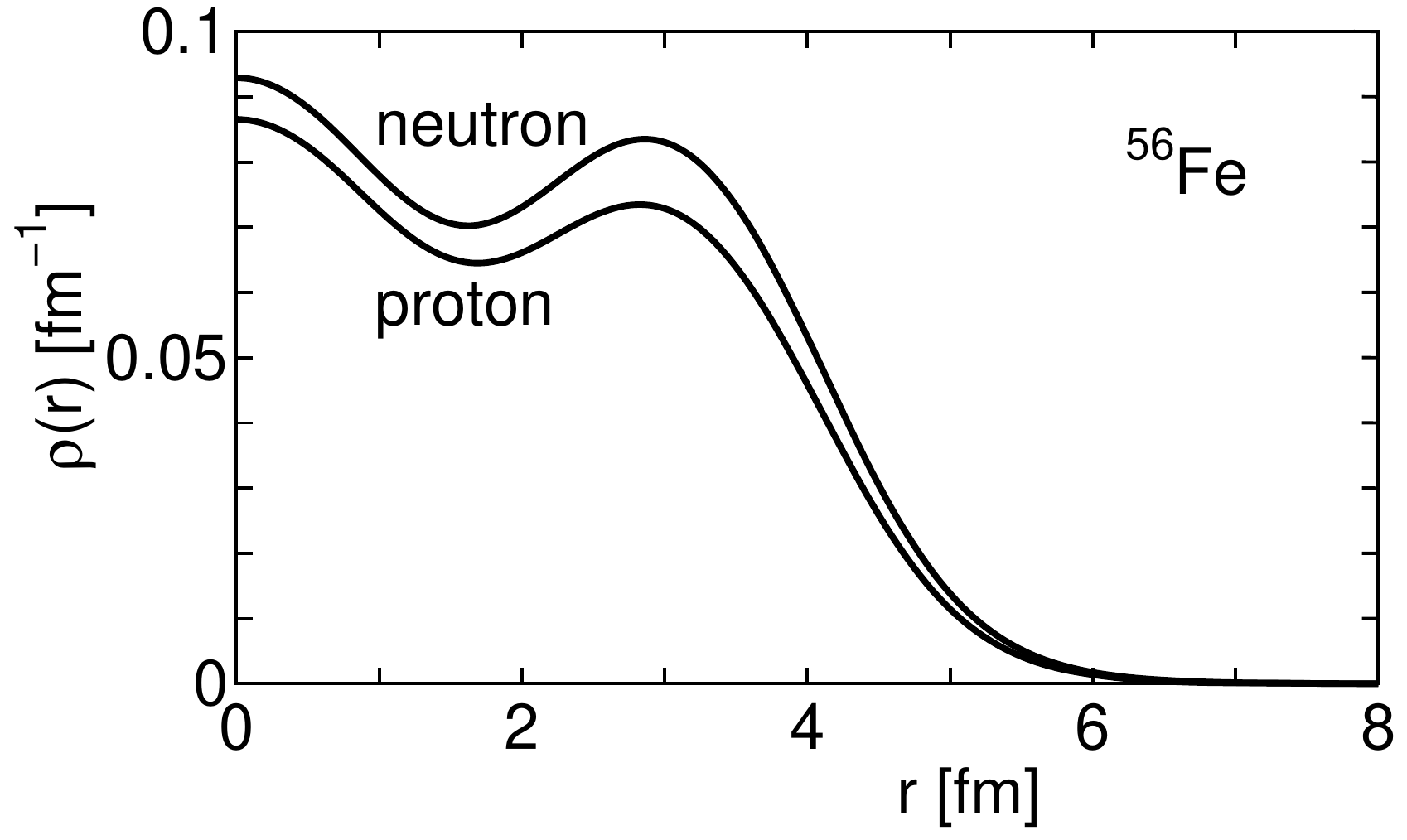}
\caption{Point proton and neutron density distributions $\rho(r)$ of $^{56}$Fe
obtained by density-dependent Hartree-Fock calculations with the
G0 force of Sprung and Banerjee \cite{SB71}. 
}
\label{fig:xidpf}
\end{figure}

In applying the ILDA method \cite{MK19} to generate the $\Xi$ single-particle
potential in $^{56}$Fe using $G$-matrices evaluated in symmetric nuclear matter, one has
to beware that the $^{56}$Fe nucleus is asymmetric in the proton and neutron density
distributions. At present, however, it is very demanding to perform Brueckner
self-consistent calculations of the $\Xi$ potential at various asymmetric nuclear
matter, because all single-particle potentials of the 8 octet baryons
(n, p, $\Lambda$, $\Sigma^-$, $\Sigma^0$, $\Sigma^+$, $\Xi^-$, and $\Xi^0$) have
to be determined self-consistently. Fortunately, the asymmetric effect can be
expected to be small as explained in the following. If the neutron and proton
contributions are individually written, the $\Xi$ potential is obtained by the sum of the
proton and neutron contributions, which is written in an abbreviated notation as
\begin{equation}
 U_{\Xi^-}=\rho_n G^{3} + \frac{1}{2} \rho_p (G^{1}+G^{3}),
\end{equation}
where $\rho_n$ and $\rho_p$ are neutron and proton density distributions,
respectively, and $G^{2T+1}$ represents the contribution of the $\Xi N$
$G$-matrix in the isospin $T$ channel. Introducing the asymmetry parameter
$\alpha$ as
$\alpha \equiv (\rho_n-\rho_p)/(\rho_n +\rho_p) \equiv (\rho_n-\rho_p)/\rho$,
the potential $U_{\Xi^-}$ is written as
\begin{equation}
 U_{\Xi^-}=\frac{1}{4}\rho (G^{1}+3G^{3})
 \left\{1+\alpha \frac{G^{3}-G^{1}}{3G^{3}+G^{1}}\right\}.
\label{eq:asym}
\end{equation}
The profile of the neutron and proton density distributions of the density-dependent
Hartree-Fock calculation with the G-0 force of Sprung and Banerjee \cite{SB71},
which is used in the present ILDA calculations of the $\Xi$ potential in $^{56}$Fe,
is shown in Fig. \ref{fig:xidpf}. The asymmetry of $^{56}$Fe is seen to be about
$\alpha \approx \frac{0.01}{0.15}=\frac{1}{15}$. The additional factor
$(G^{3}-G^{1})/(3G^{3}+G^{1})$ is smaller than $\frac{1}{3}$ as inferred from the 
properties of the $\Xi N$ interaction presented in Sec. II.
Therefore, the contribution of the second term in Eq. (\ref{eq:asym}) is estimated
to be at most 2 \% of the first term. This indicates that the estimation of the $\Xi^-$
potential based on $G$-matrices in symmetric nuclear matter is reliable.

\begin{figure}[t]
\centering
 \includegraphics[width=0.45\textwidth,pagebox=cropbox,clip]{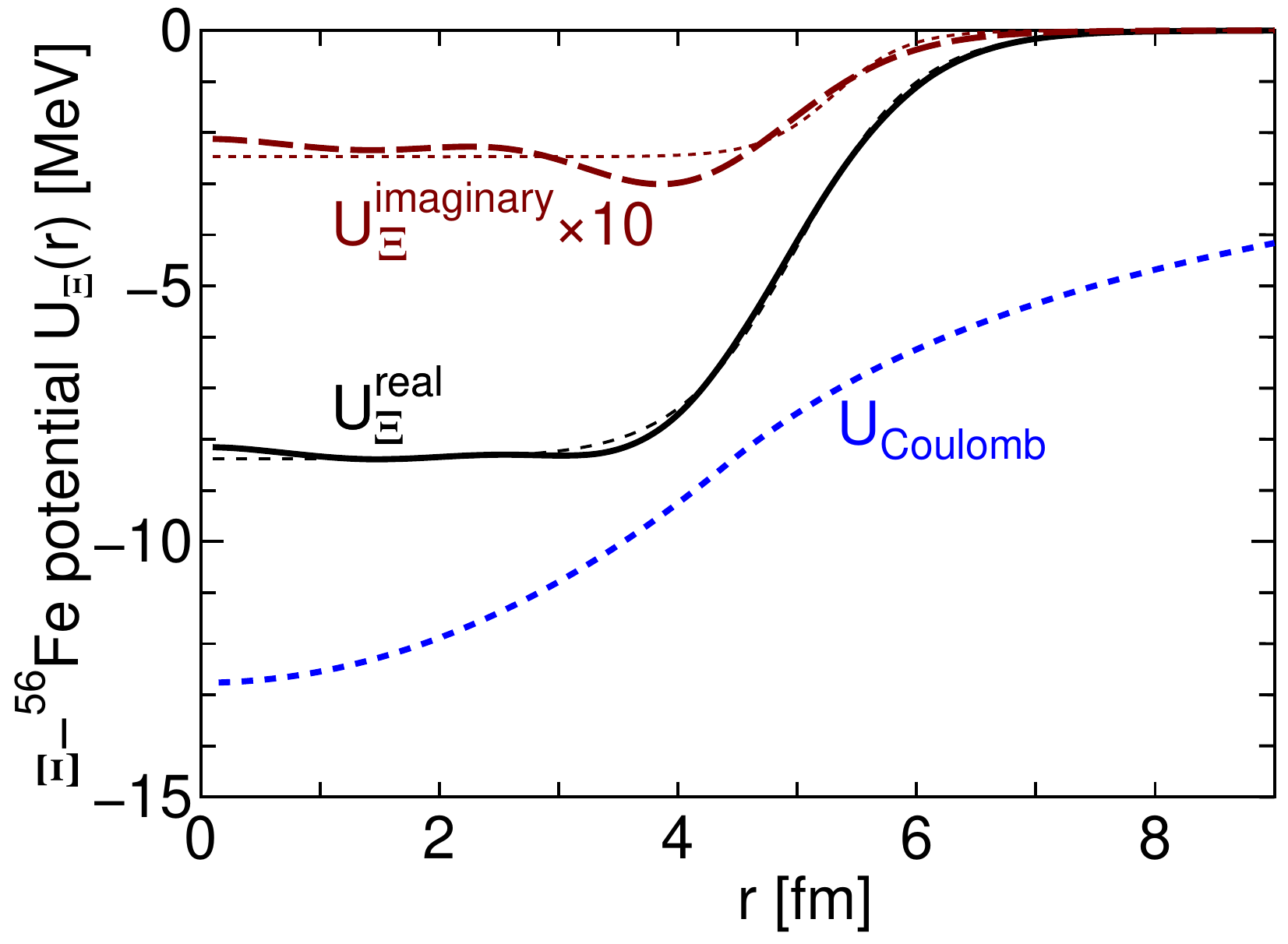}
\caption{$\Xi$ single-particle potential obtained in the ILDA method,
based on the $\Xi$ potential in symmetric nuclear matter calculated with the
NLO ChEFT interactions \cite{HAID16}. The imaginary part shown by the dashed
curve is scaled up by a factor of 10. The dotted curves are the potential fitted
in a Woods-Saxon form, the depth and the geometry parameters are given
in the text. $U_{\mbox{Coulomb}}$ depicts the Coulomb potential of the uniform
charge distribution with the radius with $R_C=1.15 A^{1/3}$ fm.
}
\label{fig:xisp}
\end{figure}

The $\Xi$-$^{56}$Fe single-particle potential calculated by the ILDA method
with the Gaussian smearing range of $\beta=1.0$ fm  is shown
in Fig. \ref{fig:xisp}. The solid and dashed curves represent
the real and imaginary parts, respectively. The potential in symmetric nuclear
matter is energy-dependent. The energy is set to be 0 MeV because the shallow
$\Xi$ level is mainly concerned to discuss the Coulomb energy level shift.
The potential shape is well simulated by a standard Woods-Saxon form both
in the real and imaginary parts. The fitted strength and geometry parameters are
$V_{R}=-8.39$ MeV, $R_{0,R}=5.01$ fm, and $a_R=0.499$ fm for the real part,
and $V_{I}=-0.247$ MeV, $R_{0,I}=5.32$ fm, and $a_I=0.303$ fm for the real part.
These Woods-Saxon potentials are shown by the dotted curves in Fig. \ref{fig:xisp}.

The depth of the real part of about 8 MeV corresponds to that in nuclear matter.
The imaginary potential, which mainly originates from the energy-conserving
$\Xi N\rightarrow \Lambda\Lambda$ process, turns out to be very small. Note
that the imaginary part is scaled up by a factor of 10 in Fig. \ref{fig:xisp}. In this
transition process, the kaon exchange has to be involved and therefore
the interaction is short-ranged. The smallness of the $\Xi N$-$\Lambda\Lambda$
coupling potential is also pointed out in the HAL-QCD calculations \cite{Sas20}.
Another factor of the smallness is the spin-isospin structure. The
$\Xi N \leftrightarrow \Lambda\Lambda$ conversion is possible only in the isospin
$T=0$ $^1S_0$ channel. The statistical factor $(2S+1)(2T+1)$ suggests that
the contribution from the $T=1$ $^1S_0$ channel is comparably suppressed,
namely 1/16 in all spin-isospin combinations of the $\Xi N$ pair.

\begin{figure}[t]
\centering
 \includegraphics[width=0.45\textwidth,clip]{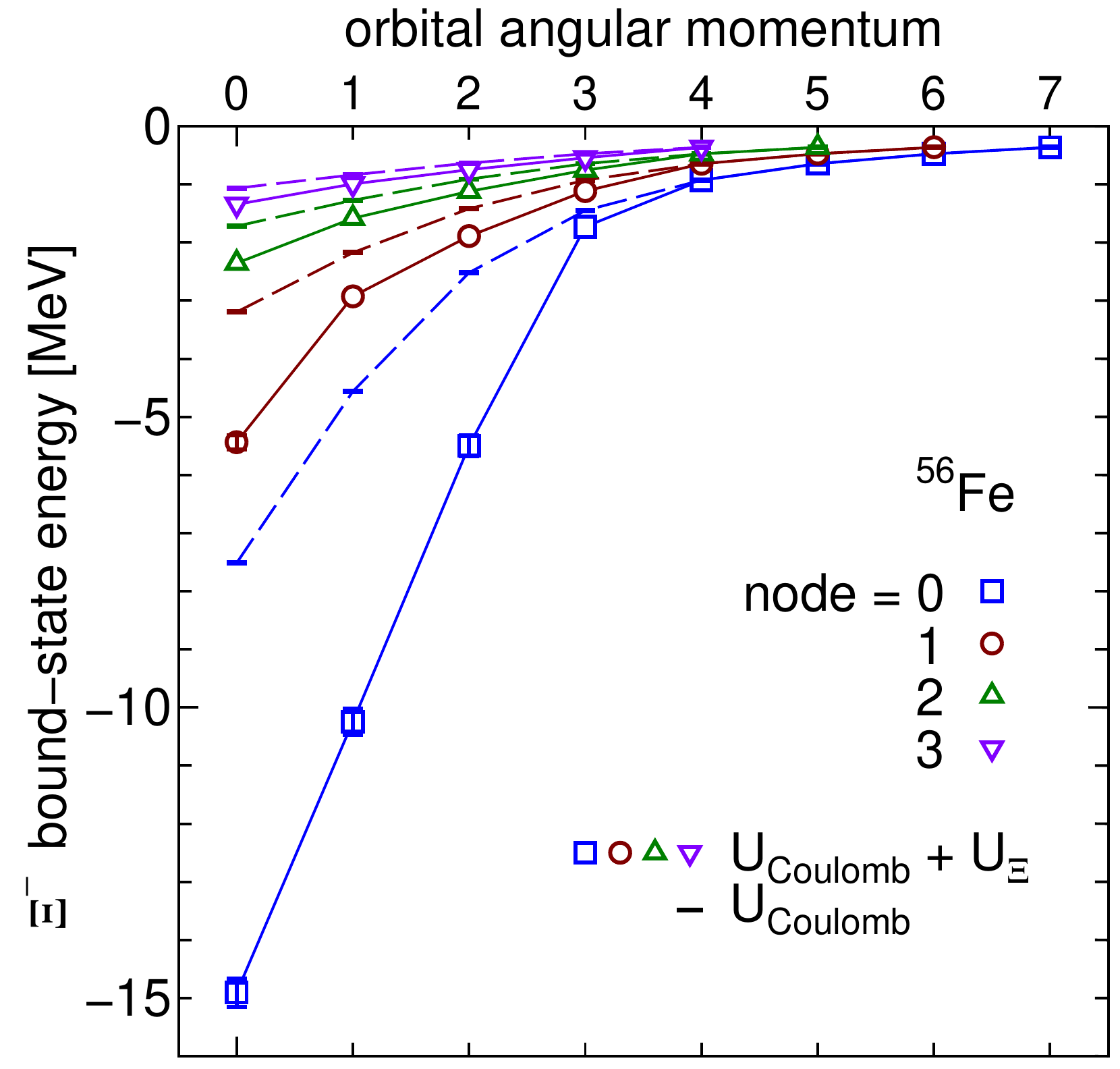}
\caption{$\Xi$ single-particle levels in $^{56}$Fe obtained by the  potential
$U_{Coulomb}+U_\Xi(E=0)$. The width $\Gamma$ is indicated
by the error bar in the case the error bar is larger than that of the symbol.
Energy levels of the pure Coulomb potential $U_{\mbox{Coulomb}}$ of the uniform
charge distribution with the radius of $R_C=1.15 A^{1/3}$ fm are also shown
with dashed connecting lines.}
\label{fig:felevel}
\end{figure}

$\Xi^-$ single-particle energies evaluated by the ILDA potential of Fig. \ref{fig:xisp}
are shown in Fig. \ref{fig:felevel}. The state is specified by its orbital angular
momentum $\ell$ and the nodal quantum number $n$.
The energies of the pure Coulomb potential of a uniform charge distribution with
a radius of $R_c=1.15 A^{1/3}$ fm are also included for comparison.
The level position $e_{real}$ and the width $\Gamma$ of the $\Xi$ state
correspond to the complex eigenvalue of the Schr\"{o}dinger equation:
\begin{equation}
 e_\Xi = e_{real}+e_{imag}=e_{real} - i\frac{\Gamma}{2}.
\end{equation}

\begin{table}[t]
 \begin{tabular}{|cc|rr|rrr|r|}\hline \hline
 node & $\ell$ & $e_C$ & $\sqrt{\langle r^2\rangle}$ & $e_{real}$ & $\Gamma$
  & $\sqrt{\langle r^2\rangle}$  & $\Delta e$ \\ \hline
  0 & 3 & $-1448.94$ &  15.3 & $-1727.81$ & $83.18$ &  10.6 & 278.87 \\
  0 & 4 &  $-927.87$ &  23.2 & $-929.66$  & $0.20$  &  23.1 &   1.78 \\
  0 & 5 &  $-644.36$ &  32.7 & $-644.37$  & $0.00$  &  32.7 &   0.01 \\
  1 & 3 &  $-927.16$ &  27.1 & $-1114.29$ & $38.52$ &  21.2 & 187.12 \\
  1 & 4 &  $-644.36$ &  37.7 & $-646.28$  & $0.24$  &  37.5 &   1.93 \\
  1 & 5 &  $-473.41$ &  49.9 & $-473.43$  & $0.00$  &  49.9 &   0.02 \\
  2 & 3 &  $-643.85$ &  41.3 & $-755.01$  & $18.26$  &  34.2 & 111.15 \\
  2 & 4 &  $-473.40$ &  54.5 & $-475.04$  & $0.20$  &  54.2 &   1.63 \\
  2 & 5 &  $-362.45$ &  69.4 & $-362.47$  & $0.00$  &  69.4 &   0.02 \\
  3 & 2 &  $-633.61$ &  44.5 & $-747.87$  & $5.30$  &  37.6 & 114.27 \\
  3 & 3 &  $-473.05$ &  57.9 & $-542.76$  & $10.16$  &  49.7 &  69.71 \\
  3 & 4 &  $-362.45$ &  73.7 & $-363.75$  & $0.16$  &  73.4 &   1.30 \\ \hline\hline
 \end{tabular}
\caption{$\Xi^-$ energies and the root-mean-square radius $\sqrt{\langle r^2\rangle}$
of the $\ell=4$ and adjacent levels evaluated with $U_{Coulomb}$ and
$U_{Coulomb}+U_{\Xi}$ in $^{56}$Fe. The level shift
given at the right end is $\Delta e\equiv e_C-e_{real}$. Entries for the
energies $e_C$, $e_{real}$, $\Gamma$, and $\Delta e$ are in keV. The unit for
the root-mean-square-radius $\sqrt{\langle r^2\rangle}$ is fm.}
\end{table}

The inclusion of $U_\Xi$ appreciably lowers the Coulomb levels with
the angular momentum $\ell \leq 3$. Reflecting the small imaginary part,
the width of the level is at most 0.5 MeV. The level with $\ell=4$ is
the main target to experimentally detect the atomic level shift by
the $\Xi$-$^{56}$Fe hyper-nuclear potential \cite{JPE03}. Table 1 tabulates energies
and predicted shifts of the $\ell=4$ and adjacent levels.
These energies are not affected by the inclusion of the $\Xi$ spin-orbit
potential argued in the preceding subsection. The
smallness of the width of the $\ell=4$ level is remarkable, though uncertainties
are kept in mind in the various stage of the present calculation.

\section{Summary}
$\Xi$ hyper-nuclear single-particle states predicted by the $\Xi$-nucleus
potential derived from the chiral NLO $\Xi N$ interactions by the
J\"{u}lich-Bonn-M\"{u}nchen group \cite{HAID16,HAID19} are presented.
To learn the basic  spin-isospin structure of the present $\Xi N$ interactions,
$\Xi N$ phase shifts are discussed and compared with those of the two sets of the
parametrization based on the HAL-QCD calculations \cite{Ino19,Sas20}.
It is also pointed out by Faddeev calculations
that no $\Xi NN$ bound state is expected in every spin-isospin state. First,
the $\Xi$ states in $^{14}$N are revisited. Considering the experimental
observation of a probable $\Xi^-$ $p$-state in $^{14}$N, the discussion is
included about the $\Xi N$ spin-orbit interactions which are relevant to the
location of the $p$-state. Then, the $\Xi$ single-particle states in $^{56}$Fe
are calculated. In particular, the atomic level shift which is expected to be measured
experimentally in the near future is predicted. The smallness of the imaginary
part of the $\Xi$ single-particle potential is demonstrated. The smallness is due
to the small transition interaction between $\Xi N$ and $\Lambda\Lambda$,
in addition to the fact that the transition to the $\Lambda\Lambda$ state
is possible only in the isospin $T=0$ $^1$S$_0$ channel.

The parametrization of the baryon-baryon interactions in the $S=-2$ sector
seems to be still in an exploratory stage due to the scarce and less-accurate
experimental scattering data.
Although $\Xi$ hyper-nuclear data is valuable, it is difficult to deduce
spin-isospin properties of the $\Xi N$ interactions by phenomenological analyses
of the experimental data of $\Xi$ states in nuclei because there are 4 spin-isospin
channels and various baryon-channel couplings are involved.
Therefore, studies based on the microscopic
baryon-baryon interactions as much reliable as possible are important.
Experimental data in the near future and theoretical microscopic studies
should improve our understanding of baryon-baryon interactions in strangeness
sectors.

\bigskip
{\it Acknowledgements.}
This work is supported by JSPS KAKENHI Grant No. JP19K03849. The authors thank
T. Inoue for supplying them with the code of the HAL-QCD potentials in the $S=-2$ sector.

\end{document}